\documentclass[10pt,
							 twocolumn,
							 superscriptaddress,
							 english,
							 prl,
							 showpacs,
							 floatfix,
							 aps,
							]{revtex4-2}
\usepackage[utf8]{inputenc}
\usepackage{amsmath}
\usepackage{amssymb}
\usepackage{graphicx}
\usepackage{xspace}
\usepackage[
separate-uncertainty  = true,
uncertainty-separator =  {\pm},
exponent-product  =  \cdot,
quotient-mode  =  fraction
]{siunitx}

\makeatletter
\@ifundefined{textcolor}{}
{%
 \definecolor{BLACK}{gray}{0}
 \definecolor{WHITE}{gray}{1}
 \definecolor{RED}{rgb}{1,0,0}
 \definecolor{GREEN}{rgb}{0,1,0}
 \definecolor{BLUE}{rgb}{0,0,1}
 \definecolor{CYAN}{cmyk}{1,0,0,0}
 \definecolor{MAGENTA}{cmyk}{0,1,0,0}
 \definecolor{YELLOW}{cmyk}{0,0,1,0}
}

\usepackage{soul}
\usepackage{braket}
\usepackage[backref=none,
bookmarksnumbered=true,
bookmarks=true,
bookmarksopen=true,
colorlinks=true,
citecolor=blue,
linkcolor=blue,
anchorcolor=green,
urlcolor=blue,unicode=false]{hyperref}

\let\baraccent=\= 
\renewcommand{\=}[1]{\stackrel{#1}{=}} 

\newcommand{\didv}{\ensuremath{\mathrm{d}I/\mathrm{d}V}\xspace}

\newcommand{\twoH}{$2H$-NbSe$_2$}
\newcommand{\nbse}{NbSe$_2$}

\makeatother

\usepackage{babel}

\newcommand{\abs}[1]{\left| #1 \right|} 

\begin{document}

\title{Wave-function engineering on superconducting substrates: Chiral Yu-Shiba-Rusinov molecules} 

\author{Lisa M. R\"{u}tten}
\affiliation{\mbox{Fachbereich Physik, Freie Universit\"at Berlin, 14195 Berlin, Germany}}

\author{Harald Schmid}
\affiliation{\mbox{Dahlem Center for Complex Quantum Systems and Fachbereich Physik, Freie Universit\"at Berlin, 14195 Berlin, Germany}}

\author{Eva Liebhaber}
\affiliation{\mbox{Fachbereich Physik, Freie Universit\"at Berlin, 14195 Berlin, Germany}}

\author{Giada Franceschi}
\affiliation{\mbox{Fachbereich Physik, Freie Universit\"at Berlin, 14195 Berlin, Germany}}

\author{Ali Yazdani}
\affiliation{\mbox{Fachbereich Physik, Freie Universit\"at Berlin, 14195 Berlin, Germany}}

\author{Ga\"el Reecht}
\affiliation{\mbox{Fachbereich Physik, Freie Universit\"at Berlin, 14195 Berlin, Germany}}

\author{Kai Rossnagel}
\affiliation{\mbox{Institut für Experimentelle und Angewandte Physik, Christian-Albrechts-Universit\"at zu Kiel, 24098 Kiel, Germany}}
\affiliation{\mbox{Ruprecht Haensel Laboratory, Deutsches Elektronen-Synchrotron DESY, 22607 Hamburg, Germany}}

\author{Felix von Oppen}
\affiliation{\mbox{Dahlem Center for Complex Quantum Systems and Fachbereich Physik, Freie Universit\"at Berlin, 14195 Berlin, Germany}}

\author{Katharina J. Franke}
\affiliation{\mbox{Fachbereich Physik, Freie Universit\"at Berlin, 14195 Berlin, Germany}}
\date{\today}

\begin{abstract}

Magnetic adatoms on superconductors give rise to Yu-Shiba-Rusinov (YSR) states that hold considerable interest for the design of topological superconductivity. Here, we show that YSR states are also an ideal platform to engineer structures with intricate wave-function symmetries. We assemble structures of iron atoms on the quasi-two-dimensional superconductor $2H$-NbSe$_2$. The Yu-Shiba-Rusinov wave functions of individual atoms extend over several nanometers enabling hybridization even at large adatom spacing. We show that the substrate can be exploited to deliberately break symmetries of the adatom structure in ways unachievable in the gas phase. We highlight this potential by designing chiral wave functions of triangular adatom structures confined within a plane. Our results significantly expand the range of interesting quantum states that can be engineered using arrays of magnetic adatoms on superconductors.
\end{abstract}

\pacs{%
			} 
\maketitle 

Structures of magnetic adsorbates on superconductors have recently garnered significant attention, primarily due to their pivotal role in the pursuit of topological superconductivity \cite{Yazdani2023}. The underlying building blocks are Yu-Shiba-Rusinov (YSR) states, which arise from the exchange interaction of magnetic adatoms with the conduction electrons of the underlying superconductor \cite{Yu1965,Shiba1968,Rusinov1969, Heinrich2018}. In his seminal work, Rusinov predicted that YSR states originating from two nearby impurities can hybridize forming symmetric and antisymmetric combinations of the YSR wave functions \cite{Rusinov1969}. This phenomenon of hybridization has been experimentally observed in both self-assembled and artificially constructed dimers \cite{Ruby2018,Kezilebieke2018,Ding2021,Kuester2021a, Beck2021}. Extended structures such as YSR chains and lattices exhibit YSR bands and have been investigated as promising platforms for topological superconductivity \cite{Kamlapure2018, Schneider2020, Schneider2021b, Mier2021, Friedrich2021, Liebhaber2022, Kuster2022, Soldini2023, Beck2023}. 

Employing magnetic atoms on superconducting substrates has the advantage of effectively isolating the induced YSR states within the gap from the bulk electronic bands. Consequently, the surface-supported electronic structures exhibit truly two-dimensional character. Alternative attempts to create adatom structures decoupled from the bulk included atomic arrangements on the surface of a semiconductor. There a two-dimensional electron gas at the surface mediates a coupling between the adatom spins while at the same time protecting them against interactions with bulk bands. In such systems, artificial molecular orbitals were realized with symmetries that are challenging to access in solution- or gas-phase chemistry \cite{Sierda2023}. 

Beyond providing electronic states that communicate a coupling among adatoms, the crystalline substrate introduces an additional interesting feature: The local symmetry of an adsorption site governs the crystal-field splitting of the states and dictates their long-range symmetry. The resulting YSR states can be used as building blocks arranged in variable symmetries on the substrate for designing larger structures. Thus, the symmetry of the substrate adds new ways to realize artificial molecules embedded in non-trivial environments, opening new opportunities for wave-function design.

Here, we use a scanning tunneling microscope (STM) to build and investigate structures of iron (Fe) atoms built atom by atom on a \twoH\ crystal. Surprisingly, we find that the spectra recorded on top of the two atoms forming a dimer differ from each other when the dimer lacks an inversion center or a perpendicular mirror plane. Such distinct spectra on individual atoms are not typically encountered in hybridized YSR dimers, emphasizing the significant influence of crystal symmetry and adsorption geometry on adatom assemblies. Employing model calculations of corresponding YSR assemblies on \nbse, we rationalize the experimental shapes and their symmetries. We exploit the symmetries of the YSR dimer to build larger two-dimensional structures and realize a versatile platform for YSR wave-function engineering as exemplified by the realization of chiral YSR molecules.


\section{YSR wave functions of $\mathrm{\textbf{Fe}}$  monomers and dimers}

\begin{figure}\centering	\includegraphics[width=\linewidth]{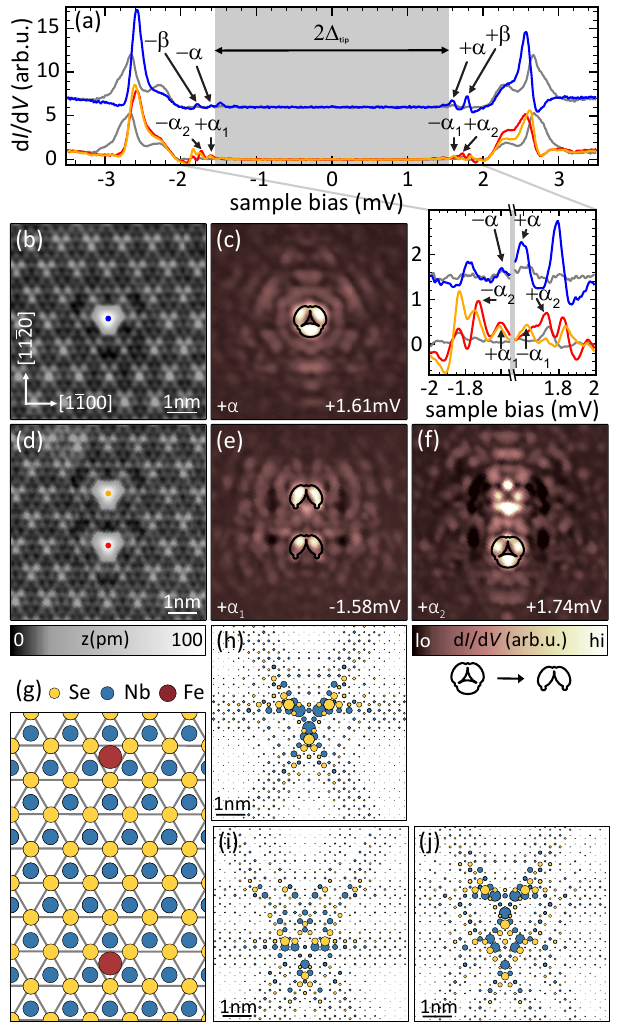}
	\caption{Fe monomer and dimer on the \twoH\ surface. (a) Tunneling spectra recorded on the Fe monomer and dimer adsorbed at CDW maxima of the \nbse\ substrate. For the color coding, refer to panels (b) and (d). Spectra of the \nbse\ substrate are shown in gray. (b), (d) Atomic resolution images of the monomer and dimer. (c), (e), (f) Corresponding constant-contour dI/dV maps of the $+\alpha$-resonance (monomer) and $+\alpha_{1,2}$-resonances (dimer) with area and scale as in (b), (c). The reduction of the characteristic $+\alpha$ shape is depicted below the color bar of (f). (g) Adsorption geometry of Fe adatoms (red circles) in the dimer configuration. (h) Numerical tight-binding calculation of the YSR monomer state (electronic density). The lattice sites are color coded as in (g) (blue for Nb and yellow for Se) and the diameter of the circles indicates the magnitude of the electronic density at the respective location.
	For parameters see Supplementary Information (SI) Note 3. 
		(i), (j) Symmetric and antisymmetric dimer wave functions. The adsorption sites are the same as in the experiment. $\Delta_{\mathrm{tip}}=$~1.55\,mV; set points: (b),(d) 10\,mV, 100\,pA; (c), top row of (a) 5\,mV, 250\,pA; bottom row of (a), (e), (f) 5\,mV, 700\,pA; all: V$_\mathrm{rms}$=15\,$\mu$V. }
	\label{fig:Fig1}
\end{figure}

When constructing structures from Fe atoms on \nbse, one has to consider the effects of the charge density wave (CDW) that coexists with superconductivity in \nbse\ at low temperatures. The YSR energy and the spatial extent and symmetry of the YSR wave functions are strongly influenced by the charge-density modulations \cite{Liebhaber2020}. To ensure that all atoms within our structures would individually exhibit equivalent spectra, we position all atoms in hollow sites of the crystal that coincide with maxima of the CDW. Figure \ref{fig:Fig1}b shows a topography of an Fe atom sitting in a hollow site at a CDW maximum (the position with respect to the lattice is indicated in Fig.\,\ref{fig:Fig1}g). The corresponding spectrum is depicted in blue in Fig.\,\ref{fig:Fig1}a with a spectrum of the bare substrate shown in gray for comparison. 
We observe four YSR states, two of which lie within the range of the substrates coherence peaks originating from the highly anisotropic band structure of \nbse\ \cite{Sanna2022}. Here, we focus on the YSR state deepest in the superconducting energy gap owing to its sharpness and characteristic spatial shape (labeled $\alpha$ in Fig.\,\ref{fig:Fig1}a).

Figure\,\ref{fig:Fig1}c shows a differential conductance (\didv) map recorded at the energy of the $+\alpha$ resonance ($+$ referring to its observation at positive bias voltage). It exhibits a distinct shape consisting of three lobes of high intensity arranged in a triangle around the atom's center. This shape was also found in previous experiments \cite{Liebhaber2020, Yang2020, Liebhaber2022}. A sketch of the shape is overlaid as a guide to the eye in Fig.\,\ref{fig:Fig1}c. Further from the atom's center, an oscillating, six-fold pattern is observed. Generally, YSR wave functions inherit their short-range characteristics from the $d$-level hosting the unpaired electron spin that gives rise to the YSR state \cite{Ruby2016}. The $d$-levels in turn are subject to a crystal field when the atom is adsorbed on a surface and therefore inherit the threefold symmetry of the adsorption geometry. Here, all atoms are adsorbed in high-symmetry positions of the substrate both with respect to the lattice and the CDW. Correspondingly, we observe $D_{3}$ symmetry (three-fold rotational symmetry including mirror axes) in both the topography (Fig.\,\ref{fig:Fig1}b) and the \didv map (Fig.\,\ref{fig:Fig1}c). The long-range oscillations of the YSR wave function reflect the hexagonal Fermi surface of the substrate with the periodicity given by the Fermi wave vector $k_F$ \cite{Menard2015, Ruby2016}. 

We form dimers by positioning a second Fe atom at a next-nearest CDW maximum. This maximum is located at a distance of $\approx$\,1.7\,nm along the [$11\bar{2}0$] direction, i.e., perpendicular to an atomic row, and corresponding to a spacing of $3\sqrt{3}a$ (with $a$ being the lattice constant). The schematic in Fig.\,\ref{fig:Fig1}g depicts the positions of the Fe atoms on the \nbse\ surface. A topographic image of the dimer is shown in Fig.\,\ref{fig:Fig1}d. Spectra recorded on both dimer atoms exhibit an increased number of YSR resonances compared to the monomer (most clearly seen in the close-up view of Fig.\,\ref{fig:Fig1}a). This larger number of resonances is a first hint towards hybridization of the YSR wave functions, so that we assign the lowest two resonances as split $\alpha$ resonances (named as $\alpha_1$ and $\alpha_2$). 

The assignment is corroborated by the \didv maps of the lowest-energy resonances. These maps show signatures of the characteristic pattern of the monomer's $+\alpha$ resonance as indicated by the overlaid sketches in Fig.\,\ref{fig:Fig1}e, f. In contrast to earlier observations on Fe dimers oriented along the [$1\bar{1}00$] direction (along the close packed Se rows), the resonances exhibit different intensities on the two atoms (compare red and orange spectra in Fig.\,\ref{fig:Fig1}a), which is also reflected in the \didv maps recorded on the dimer. Here, the $D_{3}$ symmetry is reduced to a single mirror axis. This symmetry reduction is highlighted by the change in the characteristic $+\alpha$ shape as depicted below the color bars in Fig.\,\ref{fig:Fig1}f. We note that one of the $+\alpha$ derived resonances, namely  the $+\alpha_1$ resonance, has shifted to negative bias. This transition through zero energy can be attributed to a quantum phase transition where an originally screened impurity spin with a bound quasi-particle becomes unscreened upon dimer formation. The phase transition is driven by the energy gain in Ruderman-Kittel-Kasuya-Yosida (RKKY) coupling \cite{Morr2006, Yao2014a, Schmid2022}, similar to the observation in differently arranged Fe dimers on \twoH\ \cite{Liebhaber2022}. Furthermore, the map at higher energy (Fig.\,\ref{fig:Fig1}f) exhibits characteristics of the $+\beta$ resonance alongside those of the $+\alpha$ resonance, indicating an energetic overlap of the two hybridized states (for more details see SI Note\,6). Unlike \didv maps of hybridized YSR states observed in previous experiments, neither of our $+\alpha$ maps exhibits a nodal plane between the atoms. The observation of a nodal plane for the antisymmetric hybrid state is widely considered a clear feature of hybridization \cite{Flatte2000, Morr2006, Ptok2017, Ruby2018, Liebhaber2022}. In the following, we show how this apparent contradiction is resolved.

The reduced symmetry in the YSR states of the dimer can be interpreted by simulating the spatial structure of YSR states of single atoms and assembling dimers according to the experimental geometry. We model the \nbse\ substrate by an effective tight-binding description of Nb and Se orbitals on hexagonal lattices. The Fe adatoms are placed in hollow adsorption sites of the \nbse\ lattice and treated as a classical spin impurity with isotropic exchange and potential scattering to both Nb and Se neighbors. As the Fe monomer shows long-range sixfold oscillations, we restrict the coupling to the $K$-pockets of the Fermi surface \cite{Uldemolins2022}. Different from previous models with dominant coupling to the Nb sites \cite{Menard2015}, we also include coupling to the Se sites. The simulations yield a pattern that qualitatively resembles the experimental $D_{3}$ symmetry in the YSR pattern close to the Fe atom and exhibits a six-fold symmetry in the far field (Fig.\,\ref{fig:Fig1}h). We note that our simulations do not account for the $d$-orbital structure of the adatoms, which precludes a more quantitative comparison. For details, see SI Note\,1.    

Extending the same approach to adatom dimers, we obtain the two YSR wave functions for the experimental dimer configuration as depicted in Fig.\,\ref{fig:Fig1}i, j. We observe that there is no mirror plane perpendicular to the dimer axis and the intensity distribution around the adatoms of the dimer is asymmetric. As for the monomer, the model thus correctly reproduces the symmetries observed in experiment well.

The asymmetry can also be understood within a simpler phenomenological model, which considers the YSR pattern of a single Fe atom to be a linear superposition of $s$-wave YSR states centered at the positions of the Nb atoms, i.e. at an equilateral triangle around the Fe atom (Fig.\,\ref{fig:Fig1}c, see SI Notes 2, 3). This approach is motivated by the threefold symmetry of the adatom environment so that the effective dominant exchange coupling is to neighboring Nb atoms.   
The dimer wave functions can then be simulated as symmetric and anti-symmetric linear combinations of the monomer states. This simple model captures the main symmetries found in experiment very well. In particular, our simulations are in agreement with the absence of a mirror plane perpendicular to the dimer axis. 

To further test our tight-binding and phenomenological models, we also simulated YSR wave functions of dimers with a mirror plane between both atoms (i.e. along the [$1\bar{1}00$] direction at a spacing of three lattice sites $\approx$\,1\,nm). Such dimers were probed and discussed in the context of dilute YSR chains in Ref.\ \cite{Liebhaber2022}. The models correctly reproduce the experimentally found symmetries as shown in the SI {Notes\,3 and 6}. The success of both models implies that one can readily design the symmetries of YSR wave functions in coupled adatom structures by exploiting the adsorption sites of the individual atoms and their positioning relative to each other and the underlying substrate.

\section{Chiral YSR wave functions of $\mathrm{\textbf{Fe}}$ trimers}

\begin{figure*}\centering	\includegraphics[width=\linewidth]{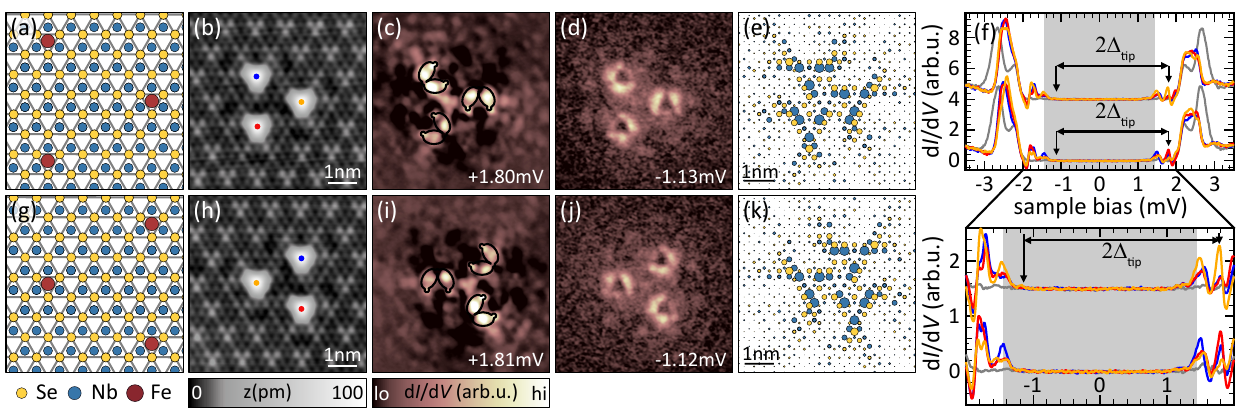}
	\caption{Chiral Fe triangles on \nbse. (a), (g) Schematic adsorption geometries of chiral Fe-trimer structures. (b), (h) Topographic images of differently oriented Fe triangles. (c), (i) Exemplary \didv map of a $+\alpha$ resonance for each enantiomer. (d), (j) \didv maps of the thermal replica of the resonances mapped in (c), (i)  (intensities multiplied by a factor of ten to compensate for the reduced intensity of thermal replica). (f) Spectra recorded on each of the atoms in the triangle depicted in (b) (top row) and (h) (bottom row). The arrows indicate the energies at which the maps were recorded. (e), (k) One YSR state of each enantiomer as obtained from our tight-binding model. $\Delta_{\mathrm{tip}}=$~1.44\,mV; set points: (b), (h) 10\,mV; 50\,pA; rest 5\,mV, 750\,pA; all: V$_{\mathrm{rms}}$=15\,$\mu$V.
	}
	\label{fig:Fig2}
\end{figure*}

We can further exploit the potential of reduced symmetry by constructing larger adatom structures in the experiment. We begin by arranging the atoms into an equilateral triangle by adding a third atom to the dimer (see Fig.\,\ref{fig:Fig2}a for a schematic of the adsorption geometry). The third atom breaks the vertical mirror symmetry of the dimer, leaving only rotational ($C_3$) symmetry. 

To probe the effect of broken symmetry on the YSR states, we first need to confirm the presence of hybridized YSR states. The corresponding \didv spectra on the individual atoms are shown in Fig.\,\ref{fig:Fig2}f. It is difficult to resolve well-separated resonances in these spectra. Yet, there is a clear change from the spectra of the monomer or dimer with an increased number of peaks. This increased number of resonances suggests further hybridization albeit with partial energetic overlap of the hybridization-derived resonances within our energy resolution. This experimental limitation may not be surprising considering the expectation of three hybrid states stemming from each of the four YSR resonances within the superconducting energy gap of only about 1\,meV. Figure\,\ref{fig:Fig2}c shows the \didv map at the energy of the sharpest peak. Interestingly, this pattern reveals a clear handedness, exhibiting only a three-fold rotational symmetry ($C_3$) without any mirror plane. Such structures are typically referred to as chiral. We identify characteristics of the $+\alpha$ derived state, which we highlight by the overlaid (reduced) characteristics. The black regions in the \didv map correspond to negative differential conductance, that occurs because we probe the energetically sharp YSR resonances with the sharp coherence peaks of the superconducting tip.

 To disentangle the actual contributions of a resonance from negative differential conductance of a close-by resonance, we look at the thermal replica of the resonance of interest. Energetically, these thermal replica are located within the energy gap of the tip (here indicated by the gray area in spectra) and shifted across zero bias by twice the superconducting gap of the tip (see SI Note\,5 for details).
A \didv map of the thermal replica of the resonance depicted in Fig.\,\ref{fig:Fig2}c is shown  in Fig.\,\ref{fig:Fig2}d. The energies at which both maps were recorded are indicated in the upper spectra in Fig.\,\ref{fig:Fig2}f. The $+\alpha$-like shape as well as the handedness are more clearly visible in Fig.\,\ref{fig:Fig2}d, but the general features are not drastically changed.
Figure\,\ref{fig:Fig2}h-j show an equivalent dataset for another trimer arrangement, i.e., where the third atom is added on the other side of the dimer (schematic of adsorption geometry in Fig.\,\ref{fig:Fig2}g). This change in configuration should produce the other enantiomer if our structure is indeed a chiral YSR molecule. The reduced $+\alpha$ shapes are mirrored in the triangle pointing to the left (Fig.\,\ref{fig:Fig2}h-j) compared to those of the triangle pointing to the right (Fig.\,\ref{fig:Fig2}b-d), revealing the opposite chirality and that the triangles are indeed enantiomers. 

Simulations of trimer states within the tight-binding model are shown in Fig.\,\ref{fig:Fig2}e and k. We have selected the equal-weight linear combination of the model monomer wave functions for both triangular arrangements.
Our simulated wave functions display a structure with reduced symmetry around each atom, essentially highlighting two sides of an equilateral triangle. The orientation of these sides obeys the threefold rotational symmetry of the trimer structure. At the same time, the rotational sense inverts from one arrangement to the other. Although the detailed shapes differ from experiment, the model correctly reproduces the observed chirality. We note that the phenomenological model also captures the chirality pattern (see SI, Note\,3).
To bring out the importance of the adsorption geometry for the appearance of chirality, we also investigate an equilateral triangle with an edge length of three lattice spacings along the [$1\bar{1}00$] direction (i.e., along the atomic Se rows). The corresponding experimental data and the modeled \didv maps are shown in the SI Note\,6. In this configuration $D_{3}$ is restored and no chirality is observed.

\begin{figure*}\centering	\includegraphics[width=\linewidth]{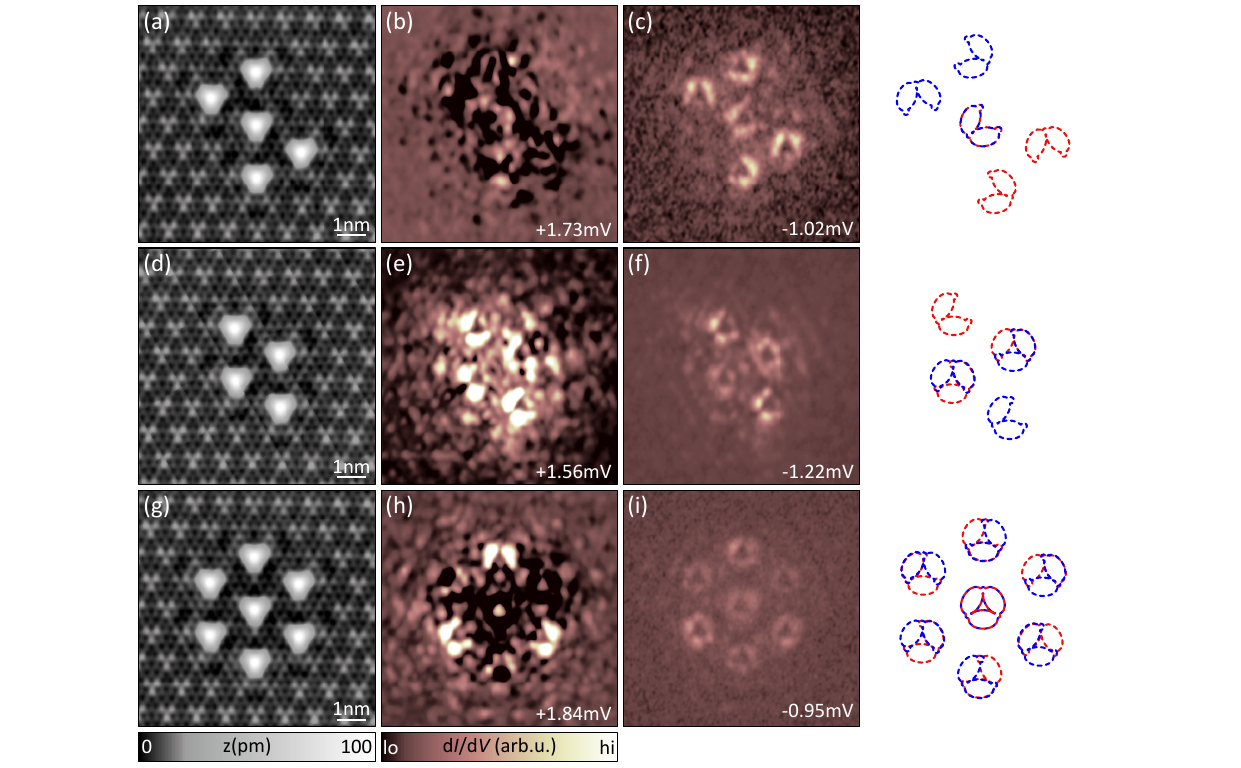}
	\caption{Larger adatom structures. (a) Topography of five Fe atoms arranged in a bowtie pattern. (b), (c) Exemplary \didv map of a $+\alpha$ resonance of the bowtie structure and its thermal replica (multiplied by a factor of ten). (d) Topographic image of four Fe atom arranged into a rhombus. (e), (f) \didv map of the rhombus structure at the energies of a $+\alpha$ resonance and its thermal replica. (g) Topography of seven Fe atoms arranged in a hexagon with an additional atom at its center. (h), (i) \didv maps of a $+\alpha$ resonance of a hexagon structure and its thermal replica (multiplied by a factor of ten to compensate for the reduced intensity of thermal replica). Next to the \didv maps we depict overlaid schematics of the reduced $+\alpha$ shapes to match the adatom structures. The different enantiomeres are indicated by different colors (blue and red). $\Delta_{\mathrm{tip}}=$~1.41\,mV; set points: (a), (d), (g) 10\,mV, 50\,pA; rest 5\,mV, 700\,pA; all: V$_\mathrm{rms}$=15\,$\mu$V.
			}
	\label{fig:Fig3}
\end{figure*}

\section{Engineering YSR wave functions in larger $\mathrm{\textbf{Fe}}$ structures}

Next, we exemplify the potential of wave-function engineering in larger structures. To this end, we deliberately tailor structures with symmetries which suppress or exhibit chiral patterns.
We start by combining two corner-sharing triangles of opposite chiralities. 
The resulting ``bow-tie'' structure is shown in Fig.\,\ref{fig:Fig3}a. We only show maps of one $+\alpha$ resonance, which features the same details as those shown for the chiral triangular YSR molecules. The \didv map of the original resonance (Fig.\,\ref{fig:Fig3}b) is strongly influenced by negative differential conductance as described above. The thermal replica shown in Fig.\,\ref{fig:Fig3}c, however, facilitates clear identification of the same patterns observed in Fig.\,\ref{fig:Fig2}c, g. The bow-tie molecule has $C_s$ symmetry, where the mirror plane coincides with the mirror plane of the reduced $+\alpha$ shape of the shared atom for each enantiomer. Therefore, the chiral pattern appears just like the one of two triangles of opposite chirality. This behavior is further visualized by a schematic of the bow tie structure, where the enantiomers are distinguished by color as depicted next to the corresponding \didv maps.

We can alternatively combine two triangles of opposite chirality resulting in a ``rhombus'' as shown in Fig.\,\ref{fig:Fig3}d. Here, two atoms are shared between the triangles and just like the bow tie, the resulting structure has $C_s$ symmetry. \didv maps of a $+\alpha$-like resonance as well as its thermal replica are shown in Fig.\,\ref{fig:Fig3}e and f, respectively. We only observe clear reduced $+\alpha$ shapes on the non-shared atoms of the structure. However, no two atoms within a chiral structure share a mirror plane. Therefore, the mirror plane of the rhombus cannot coincide with a mirror plane of both atoms that lie on it, as visualized by the schematic depicted next to the maps in Fig.\,\ref{fig:Fig2}e, f. As a consequence, the pattern observed on the shared atoms deviates from the reduced $+\alpha$ shape. 

We finally reintroduce $D_{3}$ symmetry (which we so far only observed in the monomer), by assembling six Fe atoms in an equilateral hexagon and adding a seventh atom in the center. The resulting structure is depicted in Fig.\,\ref{fig:Fig3}g. This hexagon could be reduced to any of the previously discussed structures solely by removing atoms. The \didv maps of a $+\alpha$ like resonance and its thermal replica are shown in Fig.\,\ref{fig:Fig3}h and i, respectively. When looking at chiral triangles within the hexagon all atoms are shared between triangles. Still the reduced $+\alpha$ shape can be identified in \didv maps. We can therefore trace the states of our YSR molecule back to its parent state in the monomer. 


\section{Conclusions}

In conclusion, we showed that crystalline substrates can be used for wave-function engineering of adsorbed adatom structures. This opportunity is most relevant when the wave functions originating from hybridization of the adatom states remain unperturbed from bulk states. YSR states are an ideal system as they are protected by the superconducting energy gap. Interestingly, di-atomic YSR molecules can break symmetries which cannot be broken in the gas phase. As a result, we can design complex wave-function symmetries as in chiral molecules consisting of planar arrangements of one adatom species only.
Our work focused on the design of intricate wave-function symmetries. 
Triangular magnetic adatom structures on superconductors may be even more interesting when including their spin degree of freedom \cite{Koerber2018}. Depending on the sign of the exchange interaction, the triangular structures may be ferromagnetic or frustrated. These structures offer rich opportunities for the design of chiral states with complex spin textures, eventually serving as key elements of topologically protected states. 

\section{Methods}
We achieve a clean and flat \twoH\ surface by carbon- or scotch-tape cleaving under ultra-high vacuum conditions. Fe atoms are deposited directly into the STM at temperatures below 9\,K. The as-deposited atoms are found in two distinct adsorption sites as described elsewhere \cite{Liebhaber2020}. We use superconducting Nb tips to increase the energy resolution beyond the Fermi-Dirac limit. As a consequence, all features in differential conductance (\didv) spectra are shifted by the excitation gap of the tip. Note, that the data presented here were recorded using different tips with superconducting gaps of approximately 1.55\,mV, 1.44\,mV, and 1.41\,mV (details of the tip preparation can be found in the SI Note\,5). The superconducting gap of the tip is indicated by shaded areas in all spectra.

\section{Acknowledgements}

Financial support by Deutsche Forschungsgemeinschaft through CRC 183 (project C03) and FR2726/10-1, as well as by the IMPRS ``Elementary Processes in Physical Chemistry'' is gratefully acknowledged.

%

\clearpage

\setcounter{figure}{0}
\setcounter{section}{0}
\setcounter{equation}{0}
\setcounter{table}{0}
\renewcommand{\theequation}{S\arabic{equation}}
\renewcommand{\thefigure}{S\arabic{figure}}
	\renewcommand{\thetable}{S\arabic{table}}%
	\setcounter{section}{0}
	\renewcommand{\thesection}{S\arabic{section}}%

\onecolumngrid

\maketitle 
\section*{Supplementary Information}
\section{Wave-function engineering on superconducting substrates: Chiral Yu-Shiba-Rusinov molecules}
\section{Supplementary Note 1: Tight-binding model}

We employ an effective tight-binding model to simulate the behavior of YSR molecules on \twoH. We focus on a model for the top layer neglecting the other weakly coupled layers. The top layer is itself a trilayer consisting of a central layer of Nb atoms situated between two layers of Se atoms. Both Nb and Se atoms form triangular sublattices, with the Se sublattice shifted by half a lattice vector relative to the Nb atoms. The trilayer has threefold rotation symmetry, three mirror symmetries within the plane as well as a horizontal mirror symmetry with respect to the Nb layer ($D_{3h}$ symmetry group). 

The Fermi surface consists of pockets at the $K$-points and around the $\Gamma$-point, predominantly deriving from Nb $d$-orbitals. The experimental maps of the YSR monomers show oscillations of wavelength $\lambda_F=$1\,nm, compatible with the Fermi momentum $k_F=$5.34\,nm$^{-1}$ of the $K$-pockets. This observation suggests that focusing solely on the K-pockets in our model is sufficient to capture the experimental maps. Some tight-binding models of \nbse\ in the literature \cite{SSmith1985,SRossnagel2005,SInosov2008,SMenard2015} include  only Nb orbitals, while neglecting the Se orbitals. The experimental adsorption geometry, however, suggests that exchange coupling to the Se atoms is stronger than to the Nb atoms, and may therefore be relevant.

\begin{figure}\centering	\includegraphics[width=\linewidth]{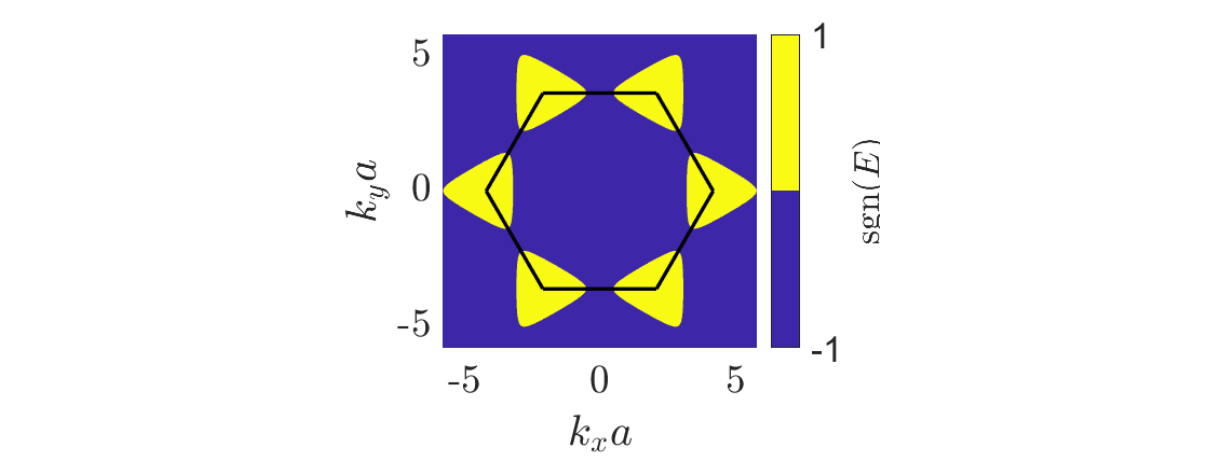}
	\caption{Fermi surface of \nbse\ as obtained from our model.
	}
	\label{fig:FS}
\end{figure}

Our minimal low-energy Hamiltonian of the normal substrate contains nearest-neighbor hopping on a hexagonal lattice of Nb and Se atoms with one orbital each,
\begin{align}
	H_0 = -t\sum_{\langle \mathbf{r},\mathbf{r}^\prime\rangle }\sum_\sigma (c^\dagger_{\mathbf{r}\sigma} d_{\mathbf{r}^\prime\sigma}+  d^\dagger_{\mathbf{r}^\prime\sigma}
	c_{\mathbf{r}\sigma})
	+V_{\mathrm{Nb}}\sum_{\mathbf{r}\sigma}c^\dagger_{\mathbf{r}\sigma} c_{\mathbf{r}\sigma}
	+V_{\mathrm{Se}}\sum_{\mathbf{r}^\prime\sigma}d^\dagger_{\mathbf{r}^\prime\sigma} d_{\mathbf{r}^\prime\sigma}
\end{align}
Here, $c^\dagger_\mathbf{r}$ ($d^\dagger_{\mathbf{r}^\prime}$) adds an electron with spin $\sigma$ to site $\mathbf{r}$ ($\mathbf{r}^\prime$) of the Nb (Se) sublattice, $t$ is the hopping amplitude and $V_{\mathrm{Nb}}$, $V_{\mathrm{Se}}$ are the sublattice potentials. The dispersion relation is
\begin{align}
	\epsilon(\mathbf{k})=\frac{V_\mathrm{Nb}+V_\mathrm{Se}}{2}
	\pm \frac{1}{2}\sqrt{(V_\mathrm{Se}-V_\mathrm{Nb})^2+4t^2\left(3+2\cos(k_y a) + 4\cos(\frac{k_ya }{2})\cos(\frac{\sqrt{3}k_xa }{2}) \right)}
\end{align}
with lattice constant $a=$~0.344\,nm. For $V_{\mathrm{Nb}}=V_{\mathrm{Se}}=0$, this expression is the low-energy description of graphene with linear dispersion at the $K$-points. For $0<V_{\mathrm{Nb}}\ll V_{\mathrm{Se}}$, the sublattice potential opens a gap and the Fermi energy (taken at zero energy) is shifted into the lower band to recover the $K$-pockets characteristic of \nbse\ with strong Nb character. 
The pockets remain disconnected as long as the shift remains small compared to the bandwidth (for $V_{\mathrm{Se}}V_{\mathrm{Nb}}<t^2$ precisely). 
Choosing $t=$1\,eV as the unit of energy, we set $V_{\mathrm{Se}}/t=4$ and $V_{\mathrm{Nb}}/t=0.22$ to match the values of the Fermi momentum $k_Fa=1.84$ from the experiment and the Fermi velocity $\hbar v_F/a=$0.78\,eV from Ref.\,\cite{SSticlet2019}. 
The Fermi surface obtained within our model is depicted in Fig.\,\ref{fig:FS}. The model fails to reproduce the $\Gamma$-pockets. It also does not reproduce the weak corrugation of the $K$-pockets along the $c$ axis, which reflects the weak interlayer coupling. We also neglect spin-orbit coupling \cite{SXi2016} and the presence of the charge density wave \cite{SBorisenko2009}. While the former splits the Fermi pockets at the $K$ points, leading to slight differences in $k_F$ for different spins, the latter respects the symmetry of the adsorption geometry. Neither effect is expected to impact the symmetry of local YSR wave functions.

The superconducting state of \nbse\ exhibits at least two distinct superconducting gaps, indicating either anisotropic or multiband pairing \cite{SSanna2022}. However, the anisotropic nature of the gap affects the YSR wave function on scales of the coherence length $\xi\simeq$~240\,nm which greatly exceeds the size of YSR molecules studied in our experiment ($d<$~3\,nm). Hence, we include superconductivity as conventional BCS s-wave pairing of equal strength $\Delta/t=0.02$ for both Nb and Se atoms, with $\xi\simeq$~20\,nm~$\gg d$. For numerical reasons, we choose $\Delta=$~20\,meV, which is an order of magnitude larger than the experimental value of $\Delta_{\mathrm{exp}}\simeq$~1\,meV. This facilitates the use of smaller lattices with $160 \times 160$ sites, and should be adequate for modeling the wave function of YSR molecules.

In our model, we treat the magnetic adatoms as classical magnetic moments. The adatoms induce exchange and potential scattering of conduction electrons. It is crucial to account for the local adsorption geometry in the experiment, where the adatoms stabilize in a $D_3$-symmetric configuration at the center of a lattice hexagon, as shown in Fig.\ \ref{fig:phenomenological}(a). Exchange and potential scattering processes of conduction electrons via $d$ orbitals of the magnetic adatom effectively couple with and between all the neighboring sites. It is essential to retain the coupling between different sites so that one obtains only a single YSR subgap state at positive energies per adatom. (In the absence of the coupling between  sites, one would obtain one YSR state per coupled site.) Our model allows for coupling to both Se and Nb atoms,
\begin{align}
	H_{I} = & \sum_{\sigma\sigma^\prime} 
 \sum_{\mathbf{e}_i\mathbf{e}_j}
 c^{\dagger}_{\mathbf{e}_i\sigma}
	\left(\frac{K_{\mathrm{Nb}}}{3}\delta_{\sigma\sigma^\prime}-\frac{J_{\mathrm{Nb}}}{3}\sigma^{z}_{\sigma\sigma^\prime}\right)
	c_{\mathbf{e}_j\sigma^\prime} +
	\sum_{\sigma\sigma^\prime}\sum_{\mathbf{e}^\prime_i\mathbf{e}^\prime_j}
 d^{\dagger}_{\mathbf{e}^\prime_i\sigma}
	\left(\frac{K_{\mathrm{Se}}}{3}\delta_{\sigma\sigma^\prime}-\frac{J_{\mathrm{Se}}}{3}\sigma^{z}_{\sigma\sigma^\prime}\right)
	d_{\mathbf{e}^\prime_j\sigma^\prime}
  \notag
\\	+&\sum_{\sigma\sigma^\prime}\left(\sum_{\mathbf{e}_i\mathbf{e}^\prime_j}
c^{\dagger}_{\mathbf{e}_i\sigma}
	\left(\frac{K_{\mathrm{Nb,Se}}}{3}\delta_{\sigma\sigma^\prime}-\frac{J_{\mathrm{Nb,Se}}}{3}\sigma^{z}_{\sigma\sigma^\prime}\right)
	d_{\mathbf{e}^\prime_j\sigma^\prime}
+ \mathrm{h.c.}\right)
 ,
	\label{eq:exchange scattering}
\end{align}
where the vectors $\mathbf{e}_i$ ($\mathbf{e}^\prime_i$) with $i=1,2,3$ locally connect the three neighboring Nb (Se) atoms to the impurity, $J_{\mathrm{Nb}}$ ($J_{\mathrm{Se}}$) is the intra-atom exchange scattering between Nb (Se) atoms, $J_{\mathrm{Nb,Se}}=\sqrt{J_{\mathrm{Nb}}J_{\mathrm{Se}}}$ the inter-atom exchange scattering from Nb to Se (and back) and $K_{\mathrm{Nb}}$, $K_{\mathrm{Se}}$ and $K_{\mathrm{Nb,Se}}=\sqrt{K_{\mathrm{Nb}}K_{\mathrm{Se}}}$ the corresponding potential scattering amplitudes. We choose $J_{\mathrm{Se}}/t=0.16$, $J_{\mathrm{Nb}}/t=0.1$, $K_{\mathrm{Nb}}/t=0.05$ and $K_{\mathrm{Se}}/t=0$. This choice reflects that the adatoms are closer to the upper Se layer than to the Nb layer.  Larger adatom structures such as those probed in experiment are treated accordingly.

\section{Supplementary Note 2: Phenomenological model}

We qualitatively compare our tight-binding simulations to a phenomenological model in which we assume that the YSR wave function can be viewed as superpositions of circular waves emanating from the three Nb neighbors of the adatom. The lattice structure enters only through the locations from which the circular waves emanate, but is otherwise effectively neglected. In particular, the phenomenological model neglects the long-range behavior of YSR states associated with the Fermi-surface geometry. Nevertheless, the phenomenological model is helpful in interpreting the symmetries of the measured YSR states as well as the YSR wave functions  obtained from the tight-binding model in the vicinity of the impurity. In our local picture, the electron (hole) wave function $\phi_+$ ($\phi_-$) emanating from the Nb atom at $\mathbf{r}_i$ has the form of a YSR state in 2D with $s$-wave symmetry \cite{SMenard2015},
\begin{align}
	\phi_{i,\pm }(\abs{\mathbf{r}-\mathbf{r}_i})=\frac{1}{\sqrt{N\pi  k_F\abs{\mathbf{r}-\mathbf{r}_i} }}
	\sin(k_F\abs{\mathbf{r}-\mathbf{r}_i}-\frac{\pi}{4}+\delta^{\pm})e^{-\sin(\delta^+-\delta^-)\abs{\mathbf{r}-\mathbf{r}_i}/\xi},
\end{align}
with the scattering phase shifts $\delta^{\pm}$ determined by potential scattering, superconducting coherence length $\xi$ and normalization constant $N$. The monomer wave function is taken to be a symmetric superposition of the circular waves emanating from the three Nb scatterers located on the vertices of an equilateral triangle,
\begin{align}    
	\psi_{M,\pm}(\mathbf{r})=\frac{1}{\sqrt{3}}\sum^3_{i=1}\phi_{i,\pm}(\abs{\mathbf{r}-\mathbf{r}_i}).
\end{align}
Dimer and trimer wave functions will be constructed by further linear combinations of monomer wave functions in the experimental configurations, assuming equal overlaps of the monomer states.

\section{Supplementary Note 3: Numerical results}

Our tight-binding model provides qualitative understanding of the experimental $dI/dV$ maps of YSR molecules. While symmetry features are reproduced, our simulations do not aim at quantitative agreement with the experimental data in view of the simplicity of the model. 

\subsection{Monomer}

In Fig.\,1(h) of the main text we show the electronic part of the simulated monomer wave functions ($|\psi|^2$) with mid-gap energy. The wave function reflects the $D_3$-symmetry of the local adsorption geometry. Notably, the maxima are directed towards the neighboring Se atoms of the Fe adatom, consistent with experiment and a density functional theory calculation \cite{SYang2020}. This feature originates from exchange coupling to the Nb atoms, and is insensitive to moderate variations in $J_{\mathrm{Nb}}/J_{\mathrm{Se}}$ given the strong Nb character of the Fermi surface. We also find that potential scattering tends to align the maxima with the Se neighbors. The direction of the wave function may be further stabilized by the charge-density wave maximum. Far from the adatom, long-range oscillations mirror the hexagonal shape of the Fermi surface.  The detailed structure of the monomer wave function in the vicinity of the impurity is largely determined by the $d$-orbital physics of the Fe adatom, limiting our calculation to qualitative symmetry features.

\begin{figure}\centering	\includegraphics[width=\linewidth]{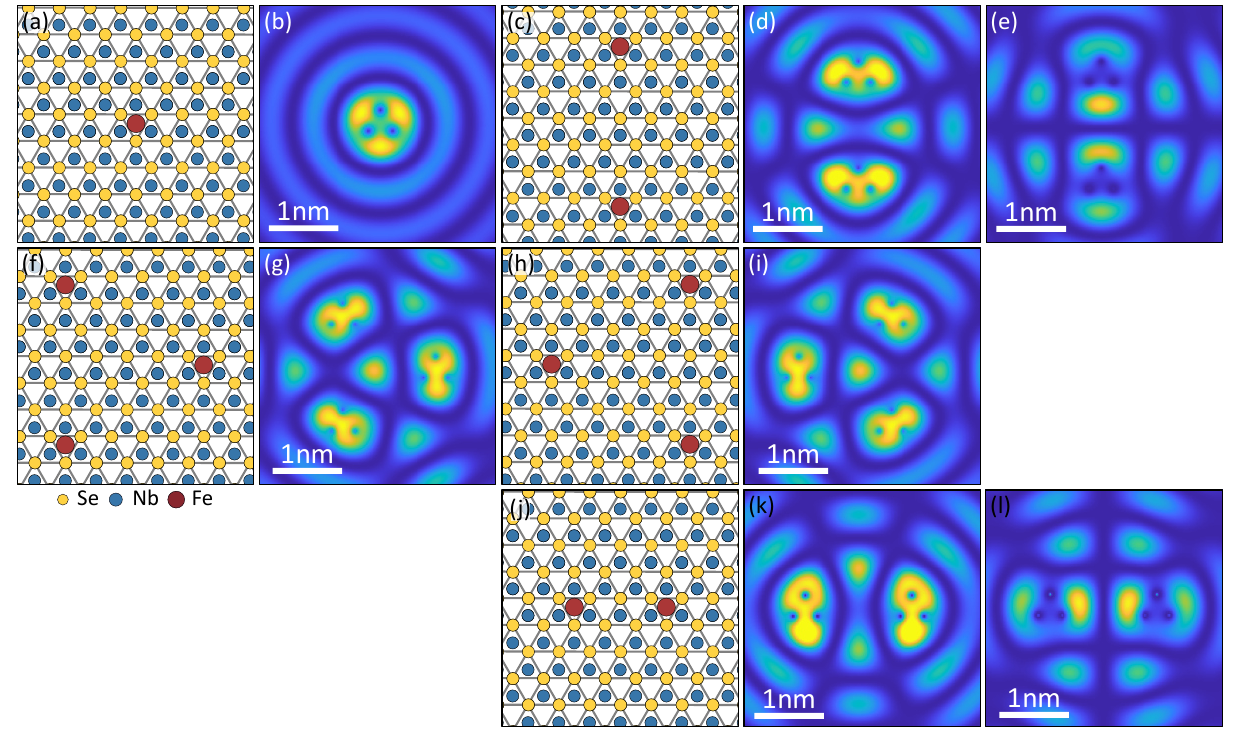}
	\caption{(a), (c), (f), (h), (j) Adsorption geometries of the monomer (a), dimer along the [$11\bar{2}0$] direction (c), trimers in configuration A (f) and B (h) formed by adding a third atom on either side of the dimer, and the dimer along the [$1\bar{1}00$] direction (j). (b), (d), (e), (g), (i), (k), (l) Wave functions obtained from the phenomenological model. Shown is the electronic part of the monomer wave function ($|\psi|^2$) (b), symmetric (d), (k) and anti-symmetric (e), (l) dimer configurations for the dimer along the [$11\bar{2}0$] (d), (e) and along the [$1\bar{1}00$] (k), (l) direction, and trimer enantiomers A (g) and B (i).
	}
	\label{fig:phenomenological}
\end{figure}

The phenomenological model gives a monomer wave function (Fig.\,\ref{fig:phenomenological}b), which is qualitatively consistent with the tight-binding calculation in the vicinity of the impurity. Further from the impurity, our ansatz shows spherically symmetric oscillations with $k_F$, in contrast to the hexagonal oscillation pattern observed in the tight-binding simulation.

\subsection{Dimers}

Fig.\,\ref{fig:phenomenological}c (and Fig.\,1g of the main text) illustrate the lattice positions of the adatoms in the dimer configuration probed in the experiment. The local adsorption geometry of the adatoms breaks the mirror symmetry with respect to a plane perpendicular to the dimer axis. Our simulations assume the same coupling for the two adatoms (Eq.\ \eqref{eq:exchange scattering}).

In Fig.\,1(i),(j) of the main text, we show the electronic part of the simulated dimer wave function (tight-binding model). The simulated wave functions reflect the broken mirror symmetry in their intensity pattern on the different adatoms, qualitatively agreeing with the experiment. The intensity patterns originate from the far-field wave fronts from one adatom interfering with the near-field wave function close to the other adatom. We have confirmed that the dimer wave functions are symmetric and antisymmetric linear combinations of the monomer wave functions. Interestingly, the antisymmetric wave function lacks a nodal plane along the broken mirror axis since the monomer wave function is not mirror-symmetric with respect to its horizontal axis. 

Comparing the tight-binding simulations with the phenomenological model (Fig.\,\ref{fig:phenomenological}b, d, e, g, i, k, l) on a qualitative level, we find that both break the mirror symmetry. However, the antisymmetric wave function of the phenomenological model displays a nodal plane, in contrast to the tight-binding model. This inconsistency arises because the phenomenological monomer wave function already exhibits approximately isotropic long-range behavior at the dimer center. In contrast, the monomer wave functions of the tight-binding model are still dominated by the threefold-symmetric short-range behavior at the dimer center, suppressing the nodal plane. Hence, the absence of the nodal plane as observed experimentally is only confirmed in the tight-binding simulation.

We also simulate dimers in a mirror-symmetric configuration (cf.\ Fig.\,\ref{fig:phenomenological} k, l and Fig.\,\ref{fig:dimer3a}i, j) placed at a distance $d^\prime=3a=$~1.03\,nm, as investigated experimentally in Ref.\,\cite{SLiebhaber2022}. The wave function of both symmetric and anti-symmetric  YSR states (Fig.\,\ref{fig:dimer3a}i, j, respectively) reflect the mirror symmetry of the adatom configuration, and exhibit a nodal plane for the anti-symmetric YSR state. The phenomenological model (Fig.\,\ref{fig:phenomenological} k, l) yields wave functions that share these properties.

\subsection{Trimers}

The trimer places a third adatom in addition to the dimer configuration, such that the three adatoms form an equilateral triangle with edge length $d$, see Fig.\,\ref{fig:phenomenological}f, h. One can place the third adatom next to the dimer in two inequivalent ways (dubbed trimer configuration $A$ and $B$) related by a mirror operation, with the original dimer axis acting as the mirror plane. Due to the adsorption geometry, the trimer has three-fold rotation symmetry, but the mirror symmetry with respect to a plane perpendicular to the edges of the adatom triangle is broken.

In our tight-binding simulation, we find three mid-gap YSR trimer states of which two are degenerate. This finding is consistent with a linear combination of the YSR monomer wave function, assuming equal hybridization between the monomer wave functions and time-reversal symmetry such that the overlap integral is real. With this ansatz, we get the trimer wave functions
\begin{align}
	\psi_{T,0}(\mathbf{r})=\sqrt{\frac{1}{3}}\bigg(\psi_{M}(\mathbf{r}-\mathbf{R}_1)+\psi_{M}(\mathbf{r}-\mathbf{R}_2)+\psi_{M}(\mathbf{r}-\mathbf{R}_3)\bigg), 
	\qquad
	\psi_{T,+/-}(\mathbf{r})=\sqrt{\frac{1}{2}}\bigg(\psi_{M}\left(\mathbf{R}-\mathbf{R}_1\right)-\psi_{M}(\mathbf{r}-\mathbf{R}_{2/3})\bigg),
\end{align}
where $\mathbf{R}_i$ are the three vectors from the trimer center to the impurities. The wave functions   $\psi_{T,+/-}$ are degenerate (so any linear combination is also possible). In the following, we focus on $\psi_{T,0}$ which is shown in Fig.\,2(h) of the main text (configuration $A$).

The trimer wave function in Fig.\,2(e) of the main text again has trilateral symmetry but no reflection symmetry. Similar to the experiment, the wave function pattern in the vicinity of each adatom describes a triangle, opening to only one of the other adatoms in a clockwise manner (positive chirality). Switching from configuration $A$ to $B$, the wave function pattern is also mirrored (Fig.\,2(k) of the main text). The opening of the triangles now is counterclockwise (negative chirality).

The chirality of the wave function pattern can be readily understood from the linear combination of monomer states. Our wave function ansatz  $\psi_{T,0}(\mathbf{r})$ is invariant under cyclic exchange of impurity vectors $\mathbf{R}_i$ reflecting the threefold rotation symmetry. However, changing the trimer configuration from  $A$ to $B$ sends $\mathbf{R}_i \rightarrow -\mathbf{R}_i$ in $\psi_{T,0}(\mathbf{r})$. Thus, the rotational direction of the  wave function pattern is reversed. The tight-binding simulation [Fig.\,2(e, k) of the main text] and the phenomenological model [Fig.\ \ref{fig:phenomenological}(g),(i)] make consistent predictions on the direction of the chiral patterns for the inequivalent enantiomers.

\section*{Supplementary Note 4: Adsorption sites and incommensurate CDW}
Single Fe atoms on the clean \nbse\ surface adsorb in two distinct sites. They can be distinguished by their different apparent height and shape and identified as sitting in two distinct hollow sites of the terminating Se layer (one with a Nb atom underneath - metal site (MS), the other one without one - hollow site (HS)) as shown in Ref.\,\cite{SLiebhaber2020}, and in the supplementary material of Ref.\,\cite{SLiebhaber2022}. Furthermore, the YSR states of Fe atoms adsorbed in hollow sites vary significantly for different adsorption sites with respect to the CDW. Details can be found in Refs.\,\cite{SLiebhaber2020} and \cite{SLiebhaber2022}. Here, we only investigate Fe atoms that sit in hollow sites on maxima of the CDW (same as atom I in Ref.\,\cite{SLiebhaber2020}, which also constituted the start of the chain in Ref.\,\cite{SLiebhaber2022}).

\section*{Supplementary Note 5: Superconducting $\mathrm{\textbf{Nb}}$ tips}
All measurements were performed with superconducting Nb tips. The tips were prepared by indenting bulk wire tips (W for the data on chiral triangles, NbTi for all other data) into a superconducting Nb sample until a sharp, stable apex and an (almost) full bulk-like superconducting gap ($\Delta\approx$1.55\,meV) was obtained. Small tip indentations were performed on the \nbse~sample to obtain a tip-apex suitable for controlled lateral manipulation of the Fe atoms. Sometimes, the superconducting gap of the tip became smaller during sample exchange or small tip formings on \nbse. Therefore, the tips used to record different data sets exhibit different superconducting gap sizes.

Because we probe the energetically sharp YSR resonances with the sharp coherence peaks of the superconducting tip, we frequently observe negative differential conductance (NDC). In \didv maps NDC manifests as black patches, where we do not resolve the actual spatial structure of a resonance.
To disentangle the contributions of a resonance from negative differential conductance, we exploit the fact that we observe thermal replica of resonances deep inside the superconducting gap. Thermal replica are only observable because we use a superconducting tip and measure at sufficiently high temperature for states close to the Fermi level to be partially occupied/ unoccupied due to Fermi-Dirac broadening. Because of this partial occupation of originally unoccupied states (and vice versa), states close to the Fermi level can be probed with both the occupied and the unoccupied coherence peak of the tip's density of states. Therefore, we observe thermal replica of those states within the energy gap of the tip (here indicated by the gray area in spectra) and shifted across zero bias by twice the superconducting gap of the tip. Thermal replica are generally less intense than the original peak and do not cause negative differential conductance.

The atoms could be moved and positioned with high precision by laterally approaching them with the STM tip in constant-current mode at set points below 10\,nA at bias voltages around 5\,mV (the exact values depend on the tip apex). We then slowly move the tip using the follow-me option in the Nanonis software and drag the atom across the surface. Each jump of the atom to a new adsorption site can be observed in the real-time current and tip-height charts. To release the atom, we change the set point to normal scanning parameters.

\section*{Supplementary Note 6: Additional data}

Here, we present additional data recorded on the structures discussed in the main manuscript as well as some complementary data on a dimer and a trimer arranged along the [$1\bar{1}00$] direction.

\subsection{Monomer and dimers}

\begin{figure}\centering	\includegraphics[width=\linewidth]{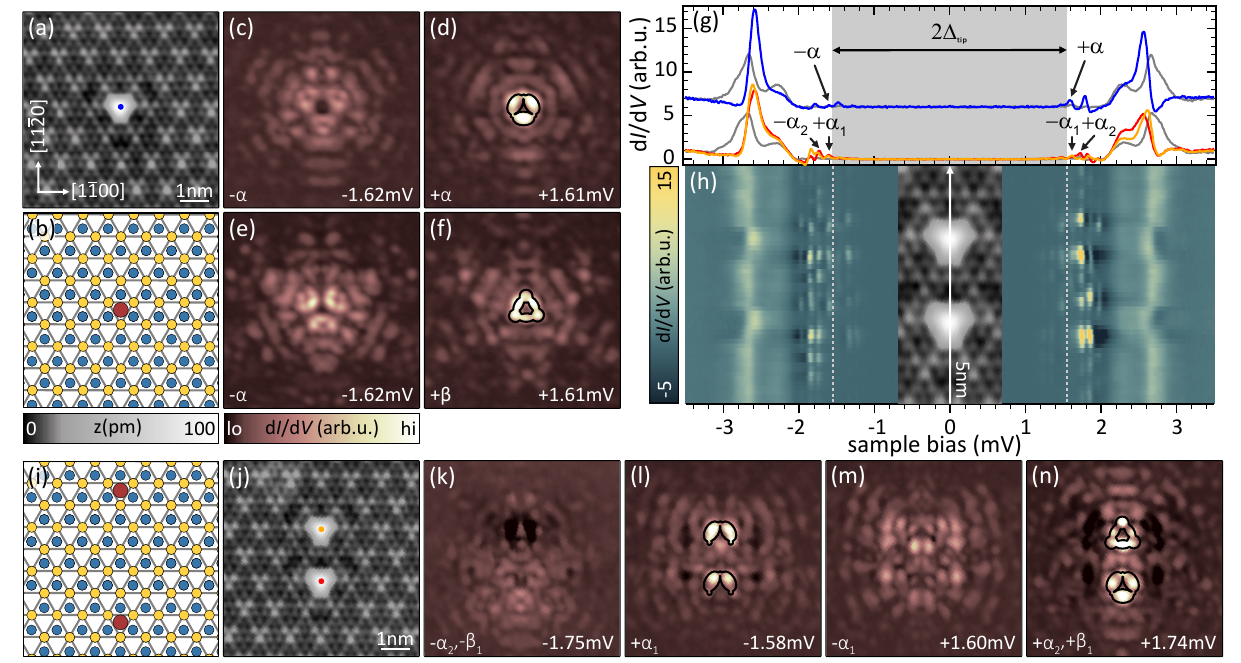}
	\caption{(a), (j) Topographic images of monomer and dimer configurations reproduced from the main text. (b), (i) Schematic adsorption geometries of the monomer (b) and the dimer along the [$11\bar{2}0$] direction (i). (c-f) \didv maps of the $\alpha$ and $\beta$ resonances of the monomer recorded at both bias polarities. (g) Spectra recorded at the positions indicated by the colored dots in the topographic images. (h) False-color plot of spectra recorded along the dimer axis (arrow depicted in the inset). (k-n) \didv maps of the two $\alpha$ resonances observed on the dimer at both bias polarities. $\Delta_{\mathrm{tip}}=$~1.55\,mV; set point: (a),(j), inset in (h) 10\,mV, 100\,pA; (c-f), top row of (g) 5\,mV, 250\,pA; bottom row of (g), (h), (k-n) 5\,mV, 700\,pA; all: V$_\mathrm{rms}$=15\,$\mu$V.
	}
	\label{fig:monomer_dimer}
\end{figure}

As mentioned in the main text, the $+\alpha_2$ map of the dimer exhibits characteristics of a monomer's $+\beta$ resonance alongside the characteristic $+\alpha$ shape. Here we explain this behavior and its origin in more detail. In Fig.\,\ref{fig:monomer_dimer}c-f we show \didv maps of the monomer's $\alpha$ and $\beta$ resonances at both bias polarities ($+\alpha$ reproduced from main text). Note that both the $\alpha$ and the $\beta$ state display distinctly different patterns at opposite bias polarities. This difference enables us to track the $+\alpha$ resonance even when it crosses the Fermi level. The characteristic $+\alpha$ and $+\beta$ shapes are overlayed as a guide to the eye in Fig.\,\ref{fig:monomer_dimer}d and f, respectively. The $+\beta$ resonance has the same overall triangular characteristics as the $+\alpha$ shape but the intensity is shifted from the edges (lobes of the $+\alpha$ resonance) to the corners, where we observe bright circles. 

We further reproduce the dimer's \didv maps shown in the main text and show their opposite polarity counter parts in Fig.\,\ref{fig:monomer_dimer}k-n. We overlay the characteristic $+\alpha$ and $+\beta$ shape in Fig.\,\ref{fig:monomer_dimer}n, which is labeled $+\alpha_2, +\beta_1$ here. Just as the $+\alpha$ shape, the $+\beta$ shape is reduced and only exhibits one mirror plane in the dimer. Further information and data on the characteristic $+\beta$ shape, its reduced form, and hybridized $\beta$ states in dimers along the [$1\bar{1}00$] can be found in previous works \cite{SLiebhaber2020, SLiebhaber2022}. Finding a signature of the $\beta$ state concomitant to the $\alpha$ state shows that the hybridization-split states become almost degenerate. 

In Fig.\,\ref{fig:monomer_dimer}h we additionally show a false-color plot of spectra recorded along the dimer axis ([$11\bar{2}0$] direction). The absence of a mirror plane between the atoms for any resonance is clearly visible in this plot. The inset shows a topography with an arrow indicating where the line of spectra was taken. In agreement with the \didv maps, all resonances that are found deep inside the gap exhibit maximal intensity next to each atoms center rather than centrally on it. Therefore the resonances appear to have very low intensities in the spectra shown in Fig.\,\ref{fig:monomer_dimer}g, which are the same spectra shown in the main text of this manuscript.

\begin{figure}\centering	\includegraphics[width=\linewidth]{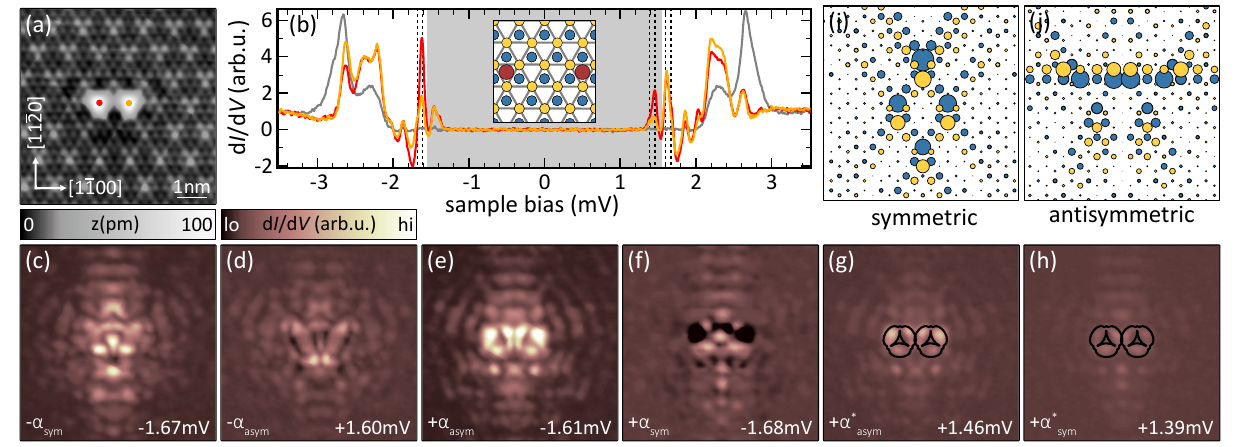}
	\caption{(a) Topographic image of a Fe dimer arranged along the [$1\bar{1}00$] direction at a distance of three lattice spacings ($\approx$\,1\,nm). (b) Spectra recorded on both atoms (positions indicated by dots of corresponding color in (a)). (c-f) \didv maps of both hybrid $\alpha$ YSR states at both bias polarities. (g, h) \didv maps of the thermal replica of the states mapped in (e, f). The energies at which all maps were recorded are indicated by dashed vertical lines in (b). (i, j) Numerical results for this dimer. $\Delta_{\mathrm{tip}}=$~1.55\,mV; set point: (a) 10\,mV, 100\,pA; rest 5\,mV, 250\,pA; all: V$_\mathrm{rms}$=15\,$\mu$V.
	}
	\label{fig:dimer3a}
\end{figure}

To highlight the symmetry properties of the dimer along [$11\bar{2}0$], we show data recorded on a dimer arranged along the [$1\bar{1}00$] direction in Fig.\,\ref{fig:dimer3a} for comparison. As visible in Fig.\,\ref{fig:dimer3a}b, both atoms exhibit equivalent spectra as expected for hybridized YSR states in dimers with a mirror plane between the atoms and also observed in previous experiments \cite{SLiebhaber2022}. Fig.\,\ref{fig:dimer3a}c-h show differential conductance maps recorded at the energies indicated by vertical dashed lines in Fig.\,\ref{fig:dimer3a}b. Note that there is a mirror plane visible between both atoms in all maps and symmetric and antisymmetric hybrid YSR states can be distinguished by a nodal line along this mirror plane. Fig.\,\ref{fig:dimer3a}g and h show thermal replica of the maps depicted in (e) and (g), respectively. A detailed analysis and discussion can be found in Ref.\,\cite{SLiebhaber2022}. We model YSR wave functions of this dimer using the same tight-binding approach discussed in Suppl.\ Note 1 and 3 as well as the main text. The resulting intensity distributions for two resonances are shown in Fig.\,\ref{fig:dimer3a}i, j. For this dimer the model yields a nodal plane perpendicular to the dimer axis for the antisymmetric combination of the monomer YSR states, which we also observe in the experiment. 

\subsection{Triangles}
\begin{figure}\centering	\includegraphics[width=\linewidth]{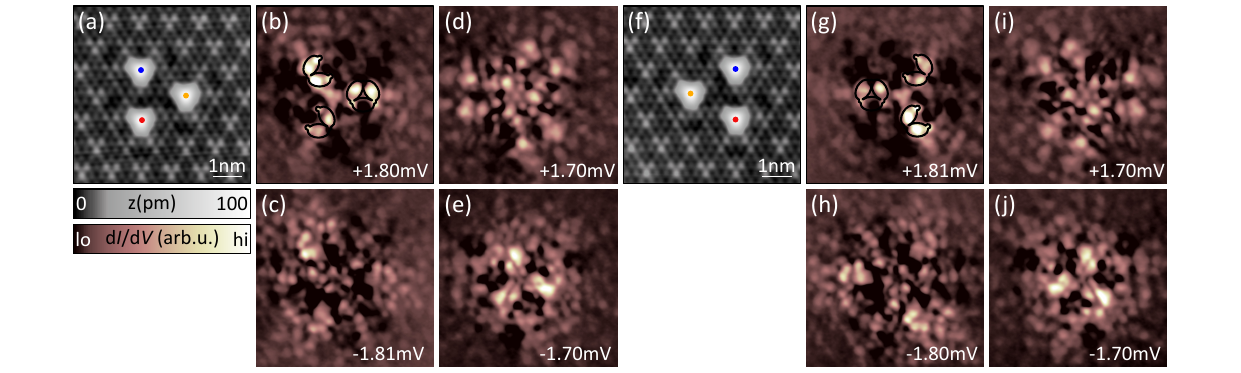}
	\caption{(a, f) Topographic images of both enantiomer triangles reproduced from the main text. (b, g) \didv maps of a $+\alpha$-like resonance of each enantiomer reproduced from the main text. (c, h) \didv map of the $\alpha$ state shown in (b, g) and the main text at opposite bias polarity. (d, e, i, j) \didv maps at both bias polarities of an additional YSR state. $\Delta_{\mathrm{tip}}=$~1.44\,mV; set point: (a), (f) 10\,mV, 50\,pA; rest 5\,mV, 750\,pA; all: V$_\mathrm{rms}$=15\,$\mu$V.
	}
	\label{fig:triangles}
\end{figure}

Fig.\,\ref{fig:triangles} reproduces data presented in the main text (a, b, f, g) alongside complimentary data(c-d, h-j). Fig.\,\ref{fig:triangles}c and h show \didv maps at the opposite bias polarity of the states shown in the main text. The chirality is less obvious in these maps than in the ones at positive bias voltages but can still be identified for example in form of a windmill-like structure at the center of each map. We further present \didv maps of another YSR state of our chiral triangles in Fig.\,\ref{fig:triangles}d, e, i, j.

\begin{figure}\centering	\includegraphics[width=\linewidth]{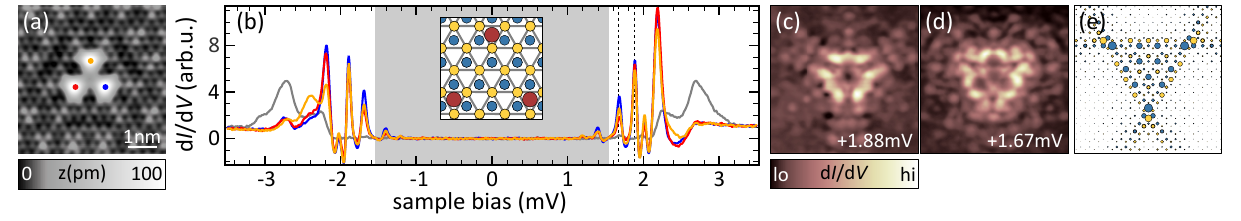}
	\caption{(a) STM topography of three Fe atoms arranged into an equilateral triangle with edges along the [$1\bar{1}00$] direction. (b) Spectra recorded on each atom of the trimer. (c, d) \didv maps of two exemplary YSR states (energies indicated by vertical dashed lines in (b)). $\Delta_{\mathrm{tip}}=$~1.55\,mV; set point: (a) 10\,mV, 50\,pA; rest 5\,mV, 250\,pA; all: V$_\mathrm{rms}$=15\,$\mu$V.
	}
	\label{fig:triangle3a}
\end{figure}

As for the dimer, we also built a triangle with sides along the [$1\bar{1}00$] direction. The corresponding data set is shown in Fig.\,\ref{fig:triangle3a}. The atoms within this structure are spaced closer to each other, than the ones in the triangles shown in the main text ($\approx$1\,nm instead of $\approx\sqrt{3}$\,nm), as visible in the topographic image shown in Fig.\,\ref{fig:triangle3a}a. Spectra recorded on each of the atoms are shown in Fig.\,\ref{fig:triangle3a}b, where vertical dashed lines indicate energies at which the \didv maps shown in c and d were recorded. The \didv maps do not show any chiral features but $C_{3v}$ symmetry. The modeled wave function captures the over all triangular shape with little intensity at the trinangles center observed in experiments.

\subsection{Beyond trimers}
\begin{figure}\centering	\includegraphics[width=\linewidth]{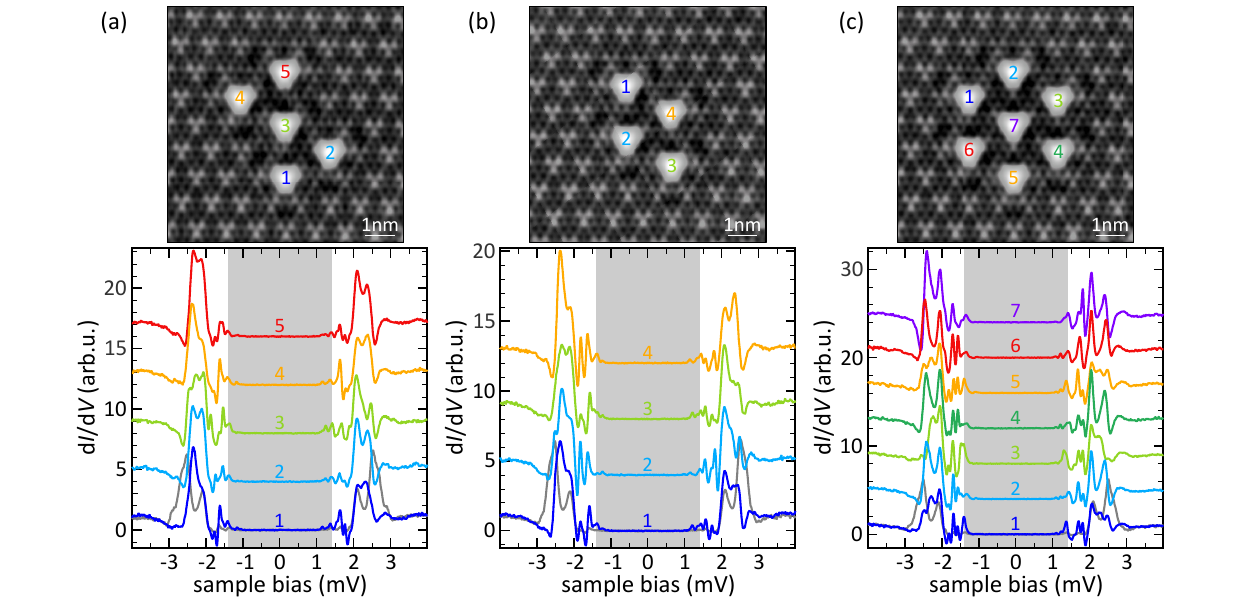}
	\caption{Spectra recorded on all atoms of the bow-tie, rhombus, and hexagon structures with the positions indicated in topographic images above. $\Delta_{\mathrm{tip}}=$~1.41\,mV; set point: topographies 10\,mV, 50\,pA; spectra 5\,mV, 700\,pA; all: V$_\mathrm{rms}$=15\,$\mu$V.
	}
	\label{fig:speclarger}
\end{figure}

Fig.\,\ref{fig:speclarger} shows spectra recorded on each atom of the bow-tie (a), rhombus (b), and hexagon (c) structures along with the same topographic images shown in the main text. The positions, at which each spectrum was recorded, are labeled and color coded in the topographies. All of the spectra exhibit several YSR states that overlap within our energy resolution.

\def\urlprefix{}
  \def\url#1.{}
%

%
\end{document}